\documentclass[aps,prl,twocolumn,superscriptaddress]{revtex4-1}

\usepackage{graphicx}
\usepackage{bm}
\usepackage{amsmath}
\usepackage{amssymb}
\usepackage{euscript}

\usepackage{verbatim}
\usepackage{fancyhdr}
\usepackage{setspace}

\usepackage{xcolor}

\usepackage{amsfonts}
\usepackage{subfigure}

\newcommand{\be}{\begin{equation}}
\newcommand{\ee}{\end{equation}}
\newcommand{\bea}{\begin{eqnarray}}
\newcommand{\eea}{\end{eqnarray}}

\begin{document}

\begin{titlepage}
	
	\title{Quantum Hall response to time-dependent strain gradients in graphene}

	\author{Eran Sela}
	\affiliation{Raymond and Beverly Sackler School of Physics and Astronomy, Tel-Aviv University, IL-69978 Tel Aviv, Israel}
	
	%\affiliation{
	%	Stewart Blusson Quantum Matter Institute, University of British Columbia, Vancouver, British Columbia, Canada}
	\author{Yakov Bloch}
	\affiliation{Raymond and Beverly Sackler School of Physics and Astronomy, Tel-Aviv University, IL-69978 Tel Aviv, Israel}
	
	\author{Felix von Oppen}
	\affiliation{Dahlem Center for Complex Quantum Systems and Fachbereich Physik, Freie Universit\"at Berlin, 14195 Berlin, Germany}
	
	\author{Moshe Ben Shalom}
	\affiliation{Raymond and Beverly Sackler School of Physics and Astronomy, Tel-Aviv University, IL-69978 Tel Aviv, Israel}

	\begin{abstract}
		Mechanical deformations of graphene induce a term in the Dirac Hamiltonian which is reminiscent of an electromagnetic vector potential. Strain gradients along particular lattice directions induce local pseudomagnetic fields and substantial energy gaps as indeed observed experimentally.
		Expanding this analogy, we propose to complement the pseudomagnetic field by a pseudoelectric field, generated by a time dependent oscillating stress applied to a graphene ribbon. The joint Hall-like response to these crossed  fields results in a strain-induced charge current along the ribbon. We analyze in detail a particular experimental implementation in the (pseudo) quantum Hall regime with weak intervalley scattering. This allows us to predict an (approximately) quantized Hall current which is unaffected  by screening due to diffusion currents. 
	\end{abstract}
	
	
	\maketitle
	
	\draft
	
	\vspace{2mm}
\end{titlepage}

Graphene offers a fertile ground to explore the rich physics of crystalline Dirac materials. A simple tight binding Hamiltonian with a constant hopping amplitude $t$ between carbon atoms gives a fair band structure description of many graphene-based systems. Famous examples include single layer graphene with its linearly dispersing (massless) Dirac fermions~\cite{wallace1947band,novoselov2005two}, electrically-biased bilayers with a displacement-field-induced band gap  ~\cite{mccann2006landau,castro2007biased}, or twisted layers with (almost) nondispersing (flat) bands and externally tunable electron correlations ~\cite{bistritzer2011moire, cao2018unconventional}.
Graphene is also outstanding in its mechanical stability. The unit cell can stretch by more than 20$\%$ without breaking~\cite{lee2008measurement}, thus allowing for significant tuning of the hopping amplitude \textit{t} by applying external stress~\cite{amorim2016novel,si2016strain,akinwande2017review}.
Combining these unique electronic and mechanical resources is highly appealing, and promises  novel ``straintronic" phenomena as well as practical electromechanical couplings.

One challenge is to open band gaps by straining the monoatomic hexagonal lattice. The two Dirac points appearing near the $K$ and $K'$ points are protected against perturbations that keep inversion and time reversal symmetries intact, as is the case for uniform (possibly anisotropic) strain. Anisotropic strain replaces the single parameter \textit{t} by three hopping amplitudes ${t_1}, {t_2}$ and ${t_3}$ [see Fig.~1(a)] and shifts the Dirac points in reciprocal space. Interestingly, the difference between the three amplitudes translates into a fictituous vector potential  $\vec A$, appearing in the Dirac Hamiltonian~\cite{manes2007symmetry},
$H= v_F \vec{\sigma} \cdot (-i \hbar \vec{\nabla} \mp e \vec A)$, 
with $eA_{x} = \frac{1}{2} \frac{\hbar}{a t } (t_2+t_3 - 2 t_1)$ and
$e A_{y} = \frac{\sqrt{3}}{2} \frac{\hbar}{a t} (t_3-t_2)$ (with
$v_F = 10^6 {\rm{m}} {\rm{s}}^{-1}$ the Fermi velocity, $a=1.4 \AA$ the lattice spacing, and $t=2.5 {\rm{eV}}$).
The strain-induced term $\vec A(\vec{r},t)$, acts within each valley as
%in a way equivalent  to 
an external electromagnetic vector potential. %$\vec A_{EM}(\vec{r},t)$. 
However, in order to preserve time reversal symmetry, this  ``pseudo" vector potential acquires opposite signs in the two valleys. Its magnitude %measures the amount of anisotropy and 
can be expressed in terms of the strain tensor components $\epsilon_{ij}$~\cite{manes2007symmetry,suzuura2002phonons,guinea2010energy},
\bea
\label{AxAy}
\vec A= \frac{\beta t}{e v_F}\begin{pmatrix} \epsilon_{xx} - \epsilon_{yy} \\ -2\epsilon_{xy} \end{pmatrix},
\eea
with $\beta =- \frac{d \log t}{d \log a} \cong 2.5$.
\begin{figure}[b]
	\centering
	\includegraphics[scale=0.4]{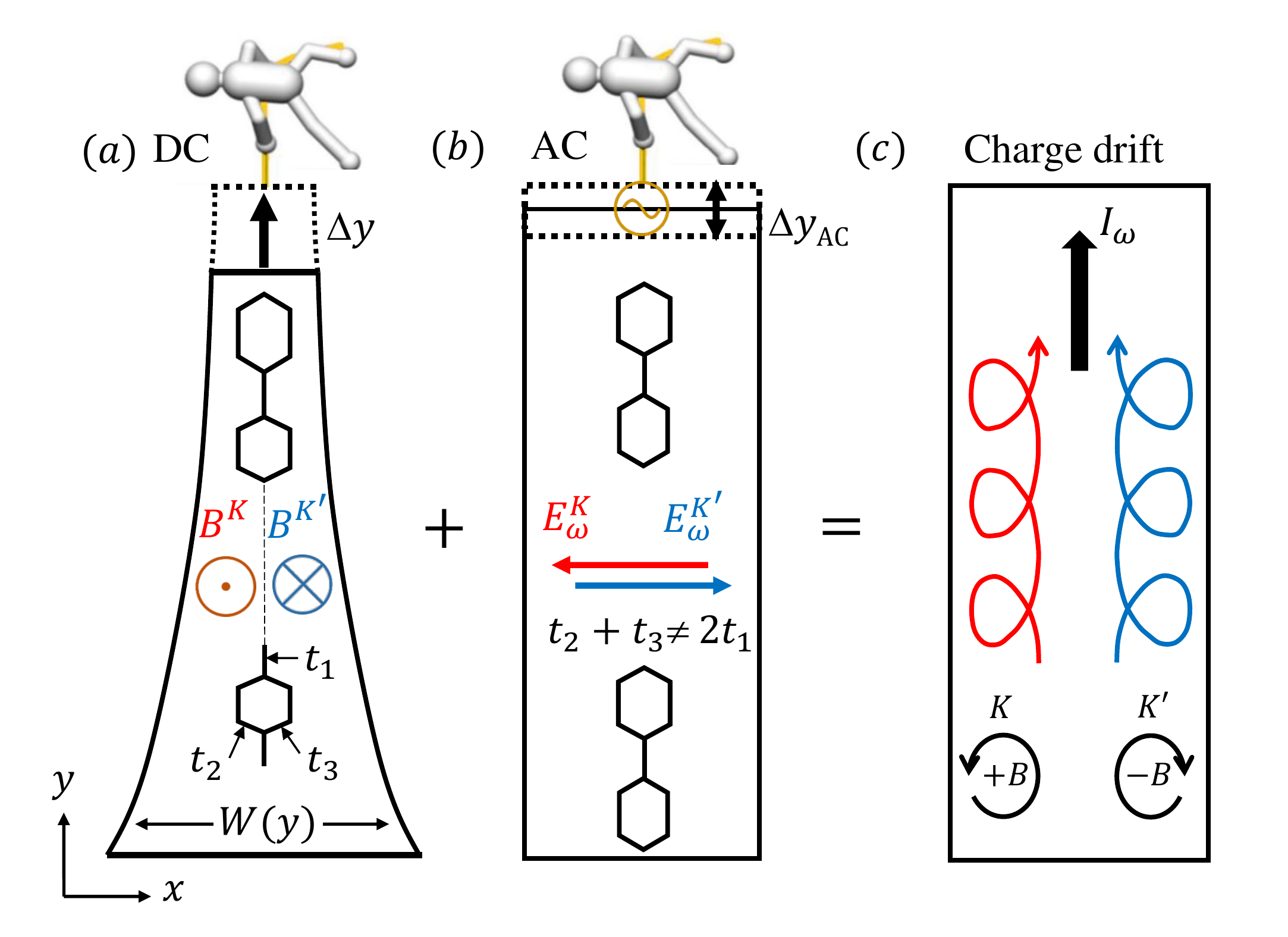}
	\caption{%Pseudo fields and electronic drift in graphene by strain engineering. 
		(a) Graphene ribbon  oriented along the armchair direction, with a uniaxial stretch generating a larger deformation at its narrow (top) end. The resulting strain gradient is designed via the shape function $W(y)$ to create a uniform pseudomagnetic field $\vec B$ (Ref.~\onlinecite{zhu2015programmable}). (b) A time varying (oscillating) stress component generates an additional pseudoelectric field $\vec E$.  The orientations of both $\vec B$ and $\vec E$ are opposite for electrons in the $K$ and $K'$ valleys. (c)  Illustration of the valley symmetric drift dynamics   considering half an oscillating period so that $\vec E$ has a fixed sign.}
	\label{fig1}
\end{figure}
For carriers in a specific valley, the pseudo vector potential $\vec A(\vec{r},t)$  implies electric and magnetic fields. The former, $\vec{E} = - \frac{d \vec A}{dt}$, is induced by time dependent strains, while the latter, $\vec B = (\vec{\nabla} \times \vec A)_z \hat{z}$, requires specific strain gradients~\cite{sasaki2005local,guinea2010energy,guinea2010generating,vozmediano2010gauge}.
Experimentally, scanning tunneling spectroscopy~\cite{levy2010strain,jiang2017visualizing} on triangularly strained graphene found %sequences of 
tunneling resonances  with a Landau-level-like spacing that indicated remarkably large pseudomagnetic fields $\vec B$ exceeding $300{\rm{T}}$.
Recent ARPES spectra~\cite{nigge2019room}  on multiple triangular islands of graphene further confirm the pseudo-Landau level (PLL) picture of flat bands separated by more than $\sim 100 {\rm{meV}}$. The %remarkably 
large energy gaps exceed room temperature and promise fascinating correlation physics as well as practical technological opportunities. 
Detecting the pseudoelectric $\vec E$ fields on the other hand remains challenging~\cite{von2009synthetic}. 
Unavoidable lattice deformations also induce a scalar potential $\phi(r,t) \propto \epsilon_{xx}(r,t) + \epsilon_{yy}(r,t)$ due to compression or dilation of the unit cell. While the scalar field acts equally on both valleys, the %pseudovector potential and hence 
pseudoelectric  field switches sign between the valleys, and %accelerates electrons from the $K$ and $K'$ valleys in opposite directions. This 
 generates valley  rather than charge currents.

Here we propose a way to observe the pseudoelectric field through charge currents by combining time-dependent and spatially varying strains which introduce both pseudo-$E$ and pseudo$-B$ fields. The concept is similar to the Hall effect where a transverse drift velocity $\vec{v}_d = \frac{\vec{B} \times \vec{E}}{|B|^2}$ is generated in the presence of non-parallel fields. While the direction of each individual field is opposite for electrons from the two valleys, the Hall-like drift velocity involves both fields and points in the same direction for both valleys.

We demonstrate this general concept in a particular and experimentally feasible geometry of a graphene ribbon under uniaxial stress ~\cite{zhu2015programmable}. As sketched in Fig.~1(a), we set the stress and the lattice armchair direction along the ribbon, while the width of the ribbon $W$ is narrowing  toward its top end. The stress leads to $t_1 \ne t_2 , t_3$ for this orientation, while the change in $W$ induces strain gradients and thus a pseudo$-B$ field. To generate a pseudo$-E$ field in the transverse direction we add a small AC stress component to the fixed DC strain, see Fig.~1(b). Together, the two intrinsic fields generate a drift motion along the ribbon for electrons from both valleys and thus an oscillating charge current $\vec{I}_{\omega} $, see Fig.~1(c). Classically, the current is given by $I_\omega = n e \frac{|E_{x}|}{|B|}W$, where $n$ is the density measured from the Dirac point. Defining the 
filling factor $\nu=\frac{h n}{eB}$, this non quantized current translates into $I_\omega = \frac{e^2}{h} \nu  V_\omega$, where $V_\omega  \sim E W$ is an AC pseudo-voltage difference induced by the intrinsic pseudo-$E$ field.
While in principle detectable, we show that $\vec{I}_{\omega}$ is usually minute  for AC strain frequencies smaller than $\sim$ GHz due to  fast and efficient screening by diffusing electrons. Our main prediction is that a sizable charge response can be observed when the pseudo-$B$ field is sufficiently strong to cause the formation of PLLs. In this regime, the presence of energy gaps efficiently suppresses screening and a Hall-like  $\vec{I}_{\omega} $ is expected for a wide range of frequencies, deformations, and doping levels, provided that intervalley scattering is weak. 
We  suggest a particular device realization where these requirements can be achieved, and compare the conventional quantum Hall (QH) response for static magnetic fields, both externally applied and strain induced, to the dynamic pseudo QH  response which is the main prediction of this paper.

\emph{System---} We envision a micrometer scale geometry as shown in Fig.~2(a). The ribbon is clamped at the top and bottom ends and pulled by metallic beams, which are also used to measure the charge transport response. We compute the mechanical response   
$\epsilon_{ij}$ by COMSOL finite element simulations, adjusting the external stretch to $\Delta y \approx 100$nm to induce a maximum local strain of 20$\%$. Beyond this stretch the narrow (top) part of the ribbon will rupture. Naturally, variations in the ribbon's width $W(y)$ translate into gradients in $\epsilon_{yy}$ 
along the $y$ direction [see Fig.~2(c)], and hence to a finite $\frac{dA_{x}}{dy}$. More specifically, the shape function $W(y)$ is selected such as to optimize a constant gradient in $\epsilon_{yy}$ and hence a uniform pseudo-$B$ field over a large section of the ribbon~\cite{zhu2015programmable,appendix}. The color map in Fig~2(b) shows $B=(\nabla \times A)_z$  and the calculated strain tensor components. For the specific dimensions presented we obtain a nearly constant $B\cong 3 $T over $0.5 \mu$m at the center of the ribbon, see the black line in Fig.~2(c).

\begin{figure}[t]
	\centering
	\includegraphics[scale=0.42]{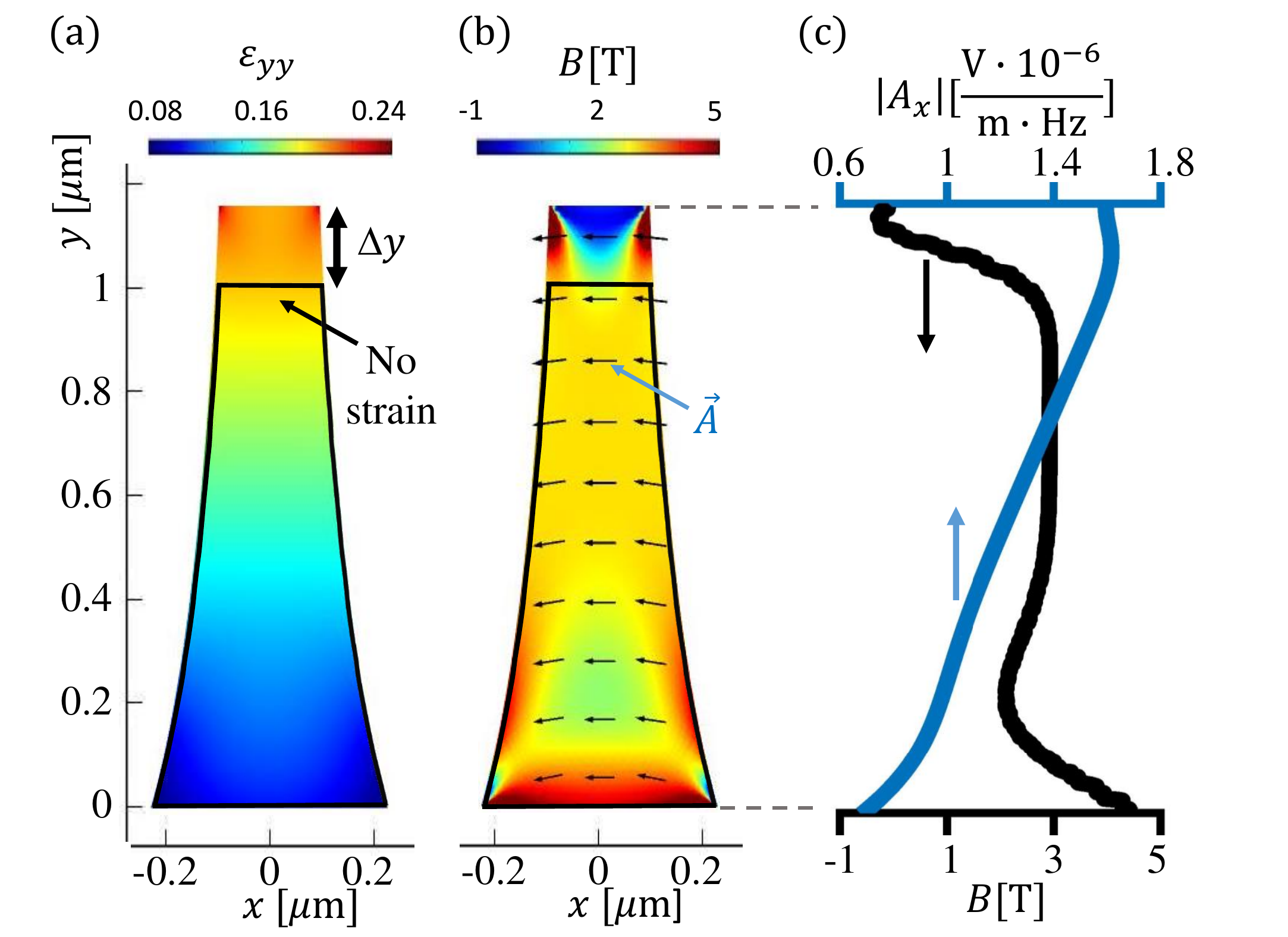}
	\caption{(a) Strain tensor component $\epsilon_{yy}(x,y)$ calculated by COMSOL. We set the vertical stretch $\Delta y$ to induce a maximum strain $\epsilon_{yy}$ of $20\%$ (red)%, which is below the breakdown strain of graphene
		. (b)  Color plot of the calculated pseudomagnetic field $(\nabla \times \vec{A})_z$,  and arrow plot of the pseudo vector potential calculated from Eq.~(\ref{AxAy}), determining the strength and direction of the pseudoelectric field for AC strain. (c) Profiles of $B$, showing a relatively uniform $\approx 3 $T over a microscale region (blue), and of $A_{x}$, allowing to extract the pseudo-voltage of $0.1$mV (see text).    }
	\label{fig2}
\end{figure}
The local pseudo-$E$ field is generated by a small AC stress applied in the same direction. 
We envision a mechanical piezoelectric manipulator that can modulate the strain at a frequency of, say, $\omega$=10MHz, and assume a small oscillation amplitude such that the device elongates by $\pm \Delta y_{AC}$ where $\Delta y_{AC} \approx 10$nm (leaving $B$ approximately unchanged). The orientation of $E$ and its magnitude (scaled by a factor  $\Delta y_{AC}\times\omega$) are determined by $A$ and presented at several points in the ribbon by the arrows (size and orientation) in Fig.~2(b). The magnitude of $A$ along the $x=0$ line is 
%also 
plotted in Fig.~2(c). As shown, the arrows are pointing primarily in the $x-$direction indicating
an AC pseudo-voltage difference between the right and left sides of the ribbon. Upon integration, $ V_\omega = \int dx E \sim W E$, we find a value $\sim 0.1$ mV.

Using a simple elastic theory one can also obtain analytic approximations of the two pseudofields ~\cite{appendix}, $E_{x}(y)=\left(\frac{\Delta y_{AC}}{L} \right) \omega \left(\frac{1}{1+f_r}\frac{W(L)}{W(y)}  \right)\left(4 (1+\bar{\nu}) \beta  \frac{t}{ev_F} \right)$, and $B= \left(\frac{\Delta y}{L} \right) \left( \frac{1-f_r}{1+f_r} \right)  \left(\frac{1}{L}  \right) \times 6 (1+\bar{\nu}) \beta \frac{t}{ev_F}$. In the latter, the three  factors display the  dependence  on the ribbon's   stretching deformation $\Delta y$,  narrowing parameter $f_r=\frac{W(L)}{W(0)}$, and overall dimension $L$. In the last dimensionful factor,  $\bar{\nu} \approx 0.17$~\cite{zhu2015programmable} is the Poisson ratio and $\frac{t}{e v_F}= 2.5 {\rm{T}} \mu$m.
The field $E_{x}$  depends directly on $\epsilon_{yy}(y)$  and thus increases along the narrowing ribbon as $1/W(y)$, see Fig.~2(c). 
Finally, we note that the frequencies considered are low compared to all relevant electronic and elastic modes and hence we will treat the pseudo-$E$ field  as quasi-static.

Both the simulated and analytic results presented above show relatively large and uniform intrinsic fields over a micrometer size sample. Compared to the nano-scale systems considered to date, this leads to several advantages: finite size effects are minute, the magnetic length $l_B=\sqrt{\frac{\hbar}{e B}}$ is significantly smaller than the  system size, and the cyclotron radius $r_c = \frac{p_F}{eB} $ can be tuned below the system's dimensions $L,W$ by an external gate. Thus, the proposed system allows us to consider the QH regime with $\omega_c \tau \gg 1$ (with the cyclotron frequency $\omega_c = v_F  / l_B$) generated by pure strain.

\emph{Quantum Hall regime: static case---} Before considering AC strain and the associated pseudo-$E$ field, we discuss transport in  the  integer QH  regime, contrasting the case of an external magnetic field $B_{\rm{ext}}$  against   an intrinsic pseudo-$B$ field, as probed by a two-probe measurement.

\begin{figure}[t]
	\centering
	\includegraphics[scale=0.39]{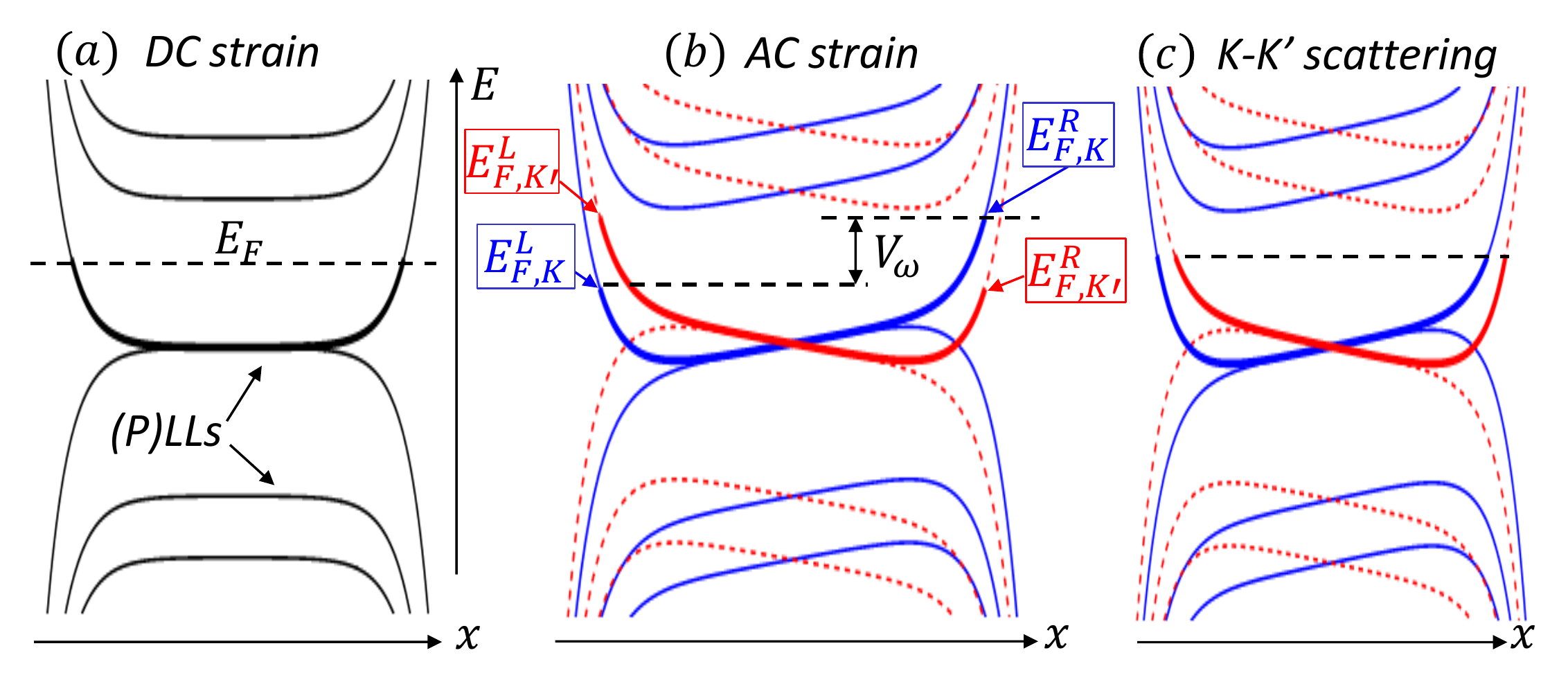}
	\caption{(a) Plot of typical PLLs (or LLs) and edge states. (b) Tilting  PLLs due to bulk pseudoelectric field caused by AC strain. A finite pseudo-voltage difference $V_\omega$ between the edges is indicated. $(c)$ Strong intervalley scattering causes equilibration within each edge.}  
	\label{fig1}
\end{figure} 

Here, we focus on the clean case, and argue that disorder just renders the QH physics   more robust~\cite{halperin1982quantized,imry2002introduction,prange1990quantum}.
To visualize the LLs and PLLs for these two cases, Fig.~3(a), consider a Corbino disk geometry (i.e., our geometry with a periodic and translation invariant $y$ direction). We can then label states by their momenta $k_y$, which are related to $x$. For an external magnetic field $k_y={\rm{const}}+ x/l_B^2$. In contrast, for a pseudo-$B$ field $k_y$ and $x$ are related differently for the two valleys, $k_y={\rm{const}}_\pm \pm x/l_B^2$.
The current follows by summing  over the contributions of all occupied states%below the Fermi  levels at the edges
, $I = \sum_{k_y {\rm{occ}}} I(k_y)$. Once the Fermi level lies 
in the bulk gap between   LLs (or PLLs), 
each (P)LL contributes a quantized current~\cite{halperin1982quantized,imry2002introduction} 
\be
\label{eq:Halperin}
I_{LL} =\frac{e}{h}  \int_{ {\rm{occ}}} dk_y \frac{d \epsilon}{dk_y} = \frac{e^2}{h}  V_{\rm{ext}}
\ee
in a two-terminal setup. Here, $V_{\rm{ext}}$ is the external voltage applied between the two terminals. 

For an external magnetic field, the edge modes are chiral. The voltage at the source terminal feeds into one of the two edges, say the right edge, elevating its chemical potential for both valleys with respect to the opposite edge~\cite{datta1997electronic}, i.e., $e V_{\rm{ext}} = E_{F,K}^R -  E_{F,K}^L = E_{F,K'}^R -  E_{F,K'}^L$. Summing over spin and valley, this  leads to a sequence of quantized plateaus in the two-terminal conductance $\frac{I}{V_{\rm{ext}}}=\frac{e^2}{h} | \nu | $, where $\nu  = \pm 2, \pm 6 \ldots$, see the black dashed curve   in Fig.~4. The current is quantized and protected by the large distance between the counterpropagating chiral edge modes. 

The pseudo-$B$ field, on the other hand, spatially superimposes counterpropagating edge states from the two valleys. Thus  the system is no longer protected against backscattering, and we expect a nonquantized two-terminal conductance~\cite{low2010strain}. The external voltage now imposes  opposite interedge chemical potential differences in the two valleys, $e V_{\rm{ext}} = E_{F,K}^R -  E_{F,K}^L =-( E_{F,K'}^R -  E_{F,K'}^L)$. Nevertheless, approximate quantization is expected if the disorder potential is smooth on the atomic scale and intervalley scattering is suppressed, see the blue curve in Fig.~4~\cite{morpurgo2006intervalley,remarksmooth}. This is possible, for example, by effectively  introducing smooth edges as described in Appendix B. %In contrast, when the sample terminates at atomically sharp edges, intervalley scattering is unavoidable, with special features emerging for zigzag edges due to their interplay with the zero PLLs~\cite{lantagne2019dispersive}.

\begin{figure}[t]
	\centering
	\includegraphics[scale=0.61]{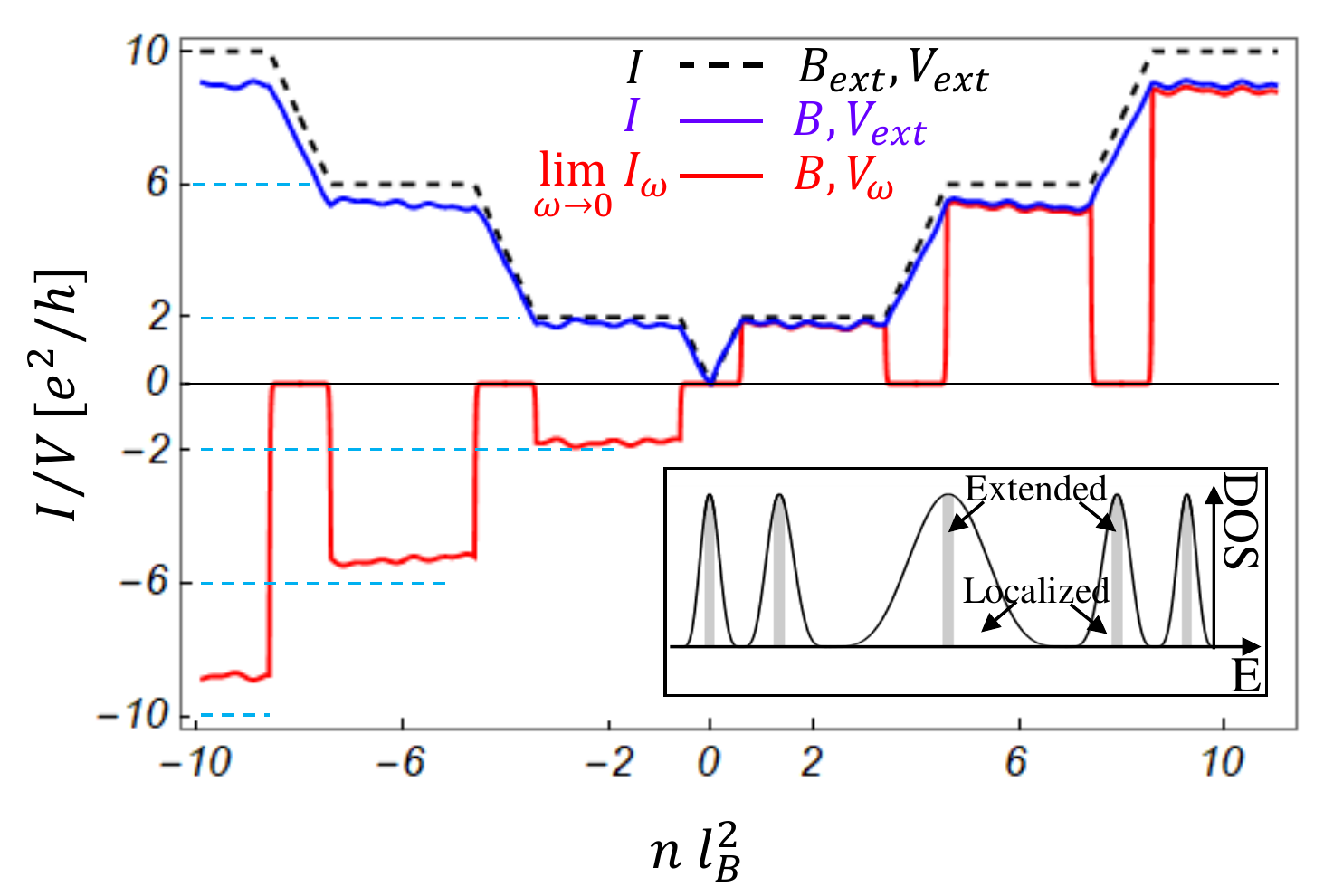}
	\caption{Schematics of  QH vs pseudo QH response as a function of density near the Dirac point. Black dashed line: two-terminal quantized conductance for external magnetic field $B_{\rm{ext}}$ and a voltage $V_{\rm{ext}}$ applied through a chemical potential difference between the terminals. Blue: two-terminal conductance for a pseudomagnetic field $B$ and  voltage $V_{\rm{ext}}$; backscattering due to counter propagating valleys leads to deviations from quantization. Red: AC current response to simultaneous pseudo-$B$ and pseudo-$E$ fields generating an internal pseudo-voltage difference $V_\omega$. The current is approximately quantized at the plateaus, and strongly suppressed at the plateau transitions due to screening of the pseudo-$E$ field~\cite{combine}. Inset: DOS consisting of extended states at the PLL energies surrounded by localized states. }  
	\label{fig4}
\end{figure} 

A few comments on our assumptions for the formation of PLLs are in order. To obtain translation invariance along the $y-$direction  we use the  gauge $A_y = -B x$ rather than $A_x = B y$. Note that our synthetic gauge theory is actually not gauge invariant in the presence of generic intervalley coupling, since the two valleys transform differently. Gauge invariance applies only to each valley separately. Exploiting  gauge invariance assumes that valley is a good quantum number. This requires negligible intervalley scattering both in the bulk and on the edges, as discussed above. Additionally, we comment that while Fig.~3(a) shows flat bands,  PLLs are not necessarily flat~\cite{lantagne2019dispersive}. %Nevertheless, the quantized conductance remains unaffected as long as the overall PLL tilt is smaller than the PLL separation. 

\emph{Quantum Hall regime: dynamic case---} Now, rather than considering a voltage difference  between  infinite reservoirs, we dynamically apply an intrinsic pseudo-$E$ field, generating a potential difference $V_\omega = E W$ between the edges. This leads to oppositely tilting PLLs in the two valleys as shown in Fig.~3(b). 
%Note that although the tilts are opposite, both valleys drift in the same direction due to their opposite relation between momentum $k_y$ and coordinate $x$.  
As long as a bulk gap remains open between the tilted PLLs, which requires $E  W < \hbar \omega_c$, the current depends solely on the chemical potential difference between the edges. We have $e V_\omega= E_{F,K}^R -  E_{F,K}^L =-( E_{F,K'}^R -  E_{F,K'}^L)$, so that Eq.~(\ref{eq:Halperin}) predicts approximately quantized plateaus for weak intervalley scattering (cf. the two-terminal case with pseudo-$B$ field). 
Thus,  in the gap-dominated regime  our AC pseudo Hall effect is quantized $
\frac{I_\omega}{V_\omega} = \frac{e^2}{h} \nu$ with $\nu=4(N+1/2)$, even at low frequencies. This AC charge current in response to the pseudo-$E$ field is our main result.
The resulting plateaus are schematically displayed in Fig.~4 by the red curve. In the presence of intervalley scattering we expect the edges to equilibrate,  see Fig.~3(c). From the above estimate of a 0.1mV pseudo-voltage, we obtain % from Eq.~(\ref{Idrift}) 
$I_{\omega}  \sim 8 {\rm{nA}}$ for filling factor $\nu=2$.% corresponding to a density of $n \sim  10^{11} {\rm{cm}}^{-2}$ at $3$T.

Notice that while the two-terminal conductance is necessarily positive, the pseudo-$E$ response in fact changes sign across the Dirac point, see the red curve in Fig.~4. This is a consequence of the sign change in the group velocity of the edge states upon going from electron to hole doping.

Another crucial difference between the %effects of an 
external voltage $V_{\rm{ext}}$ between infinite reservoirs and the intrinsically generated transverse voltage $V_\omega$ is that the latter  will be screened by electronic diffusion across the ribbon on time scales much faster than the AC frequency. This is reflected in the red curve in Fig.~4 where the current response drops to zero at QH transitions due to screening in these delocalized situations. The same  effect of the pseudo-$E$ field will occur for any gapless system, in particular for small pseudo-$B$ fields with  $\omega_c \tau \ll 1$. We now discuss this screening   at finite frequency within a semiclassical treatment.

\emph{Gapless (screened) regime---}  The interplay of a dissipative conductivity $\sigma$ with the edges of the sample and the resulting valley polarization can be described via the transport equation 
\bea
\label{eqdrudedifff}
\vec{j}_\pm = \pm \sigma \vec{E}_\omega - D \vec{\nabla} n_\pm \mp (\omega_c \tau) \vec{j}_\pm \times \hat{z},
\eea
for the current densities $\vec{j}_\pm(x,y)$ and charge densities $n_\pm(x,y)$ of the two valleys. The  conductivity $\sigma$ is related to the diffusion coefficient $D$  via the Einstein relation. The last term in Eq.~(\ref{eqdrudedifff}) is a  Hall term. The edges of the sample imply the boundary condition $j_x(x=0,y)=j_x(x=W,y)=0$. We provide a closed form solution of this equation (combined with the  continuity equation) in Appendix C. It can be written in terms of two dimensionless parameters. In addition to  $\omega_c \tau$, the typical time scale for traversing the sample, $\tau_T = W^2/D$, introduces a second dimensionless parameter $\omega \tau_T$, which controls the reduction of the current due to screening by valley-dependent diffusion  currents. 

For $\omega \tau_T \gg 1$  screening is not effective. In this regime the current takes the Drude form $\frac{I_\omega}{V_\omega}= \frac{\sigma  (\omega_c \tau) }{1+(\omega_c \tau)^2}$ for an infinite system, and is in phase with the pseudo-$E$ field. At low frequencies, we find~\cite{appendix} 
$\frac{I_\omega}{V_\omega}=\frac{i}{12} \sigma ( \omega \tau_T) (\omega_c \tau)$,
which is out of phase with the pseudo-$E$ field. %Also since $E_\omega \propto \omega$, this has a $\omega^2$ dependence on frequency. 
In the system in Fig.~1, with $\omega = 10$MHz, we estimate $\omega \tau_T \approx 10^{-3}$, leading to a strong reduction of the current at the QH transitions. For $\omega_c \tau \approx 1$ %corresponding to the quantum Hall regime, 
we obtain $I \sim p {\rm{A}} $. In principle, however, the effect can be observed even in a gapless regime, provided the frequency is sufficiently high, $\omega \tau_T \gtrsim 1$.

\emph{Conclusion---} We presented a novel mechanism to generate charge currents from space and time dependent strain fields in graphene, by combining crossed pseudo-$B$ and pseudo-$E$ fields. 
A related charge current response 
	%to a pseudo-$E$ field 
	was found~\cite{vaezi2013topological}  by gapping out graphene by a mass corresponding to a sublattice potential (e.g. due to h-BN encapsulation), which plays the role of the time reversal invariant and tunable PLL gap in our case. %While the underlying effective theory is similar, our approach may be more straight forward, general, and more importantly tunable. 
The charge current response 
should be contrasted with  previous theoretical works on transport in strained graphene, predicting valley-polarized currents in the presence of external magnetic fields~\cite{low2010strain,settnes2016quantum,ma2016mechanical,milovanovic2016strained,settnes2017valley,munoz2017analytic}, including valley filters or switches~\cite{zhai2010magnetic,fujita2010valley,wu2011valley,settnes2016graphene,stegmann2018current,prabhakar2019valley,wu2017quantum}, by combination with polarized light~\cite{golub2011valley}, parametric pumping~\cite{jiang2013generation,wang2014quantum}, or even in equilibrium in a zigzag graphene ribbon~\cite{lantagne2019dispersive}.

The relatively simple and analytic gauge field treatment of  long-wavelength strain in graphene allowed us to analyze inversion symmetry breaking on macroscopic scales, design a tunable static strain-induced bulk gap, and use these to induce a dynamic charge current. We expect this general concept to extend to a wider set of systems and materials, including %bilayer graphene or 
(3D) Weyl semimetals, which show similar synthetic gauge field effects~\cite{pikulin2016chiral,grushin2016inhomogeneous,gorbar2017chiral,arjona2018rotational}. The proposed effect also allows for measuring the edge contribution to intervalley scattering, to which it is highly sensitive even in the static case.

{\it Acknowledgements:} We thank Roman Mints for useful discussions.
%Marcel Franz, Tami Pereg-Barnea and. 
This research was supported by the Binational Science Foundation Grant No. 2016255 (ES), CRC 183 of the Deutsche Forschungsgemeinschaft (FvO), and the
Israel Science Foundation Grant No. 1652/18 (MBS).

\bibliographystyle{apsrev4-1}

%\bibliography{refs}

\begin{thebibliography}{52}
\bibitem [{rem()}]{remarksmooth}%
	\BibitemOpen
	\href@noop {} {}\bibinfo {note} {One may imagine adding smoothening backgates
		parallel to the edges of the narrowing device in order to push the electrons
		away from the disordered edge of the graphene ribbon, see Appendix
		B.}\BibitemShut {Stop}%
\bibitem [{com()}]{combine}%
	\BibitemOpen
	\href@noop {} {}\bibinfo {note} {One can also combine the dynamical pseudo AC
		voltage $V_\omega$ with an external magnetic field $B_{\rm{ext}}$. This
		generates an AC valley current instead of a charge current, which would not
		be sensitive to intervalley scattering.}\BibitemShut {Stop}%
\end{thebibliography}

%%%%%%%%%%%%%%%%%%%%%%%


%%%%%%%%%%%%%%%%%%%%%%%%

\newpage

\appendix

\section{Appendix A: Programmable gauge fields}
\label{apendix0}
\label{se:prog}
In this appendix we briefly review the relation between strain fields and synthetic gauge fields in  the geometry of  Zhu \emph{et. al.}~\cite{zhu2015programmable}. 
For completeness we review their main ideas and obtain some further useful formulas specifically for the pseudoelectric field.

Consider applying a force $F$ along the $y$ direction in Fig.~1. It leads to a stretch by $\Delta y$.  Force balance along any cut at constant $y$ implies $F=W h Y \epsilon_{yy}$ where $\sigma_{yy}= Y \epsilon_{yy}$ is the stress along $y$, $h$ is the ``width" of graphene, and $Y$ is the Young modulus. Using this simple relation one obtains a $y$-dependent strain controlled by the width function $W(y)$, \be
\label{epsilonyy}
\epsilon_{yy}(y) =\frac{F}{h Y} \frac{1}{W(y)}.
\ee Thus, a narrowing width yields a strain gradient \be
\frac{\partial \epsilon_{yy}}{\partial y} = -\frac{F}{h Y} \frac{\frac{\partial W(y)}{\partial y}}{W(y)^2}.
\ee
In order to obtain a constant gradient $\frac{\partial \epsilon_{yy}}{\partial y} $ one needs to choose a specific width function.  The specific shape function is 
\be
\label{shapefunc}W(y) = \frac{f_r L}{f_r (L-y)+y} W(0),
\ee
where $f_r=\frac{W(L)}{W(0)}$.%, one obtains $\frac{\partial \epsilon_{yy}}{\partial y} ={\rm{const}}$.

One can relate the force and the stretch $\Delta y$. Using $\Delta y = \int_0^L \epsilon_{yy}(y) dy$, and  Eqs.~(\ref{epsilonyy}) and (\ref{shapefunc}), we have 
\be
\label{ForceStretch}
\Delta y =L \frac{F}{h Y} \frac{1+f_r}{2 f_r W(0)}.
\ee

Next we would like to obtain all the strain tensor components in order to calculate the pseudo vector potential from Eq.~(1). We use the constitutive relations $\sigma_{xx} = \frac{Y}{1-\bar{\nu}^2} (\epsilon_{xx}+\bar{\nu} \epsilon_{yy})$, $\sigma_{yy} = \frac{Y}{1-\bar{\nu}^2} (\epsilon_{yy}+ \bar{\nu} \epsilon_{xx})$, and $\sigma_{xy} = 2 G \epsilon_{xy}$, as well as stress equilibrium  $\sum_{i=x,y }\partial_i \sigma_{ij}=0$. Here $\bar{\nu}$ is the Poisson ratio, and $G=\frac{E}{2(1+\bar{\nu})}$ the shear modulus. 
Assuming uniaxial stretch we have
\be
\epsilon_{xx}+\bar{\nu} \epsilon_{yy}=0.
\ee
Combining these relations one obtains~\cite{zhu2015programmable} $\frac{\partial \epsilon_{xy}}{\partial y}=0$ and $\frac{\partial \epsilon_{xy}}{\partial x}=-(1+\bar{\nu}) \frac{\partial \epsilon_{yy}}{\partial y}$. Under the condition of $\frac{\partial \epsilon_{yy}}{\partial y} ={\rm{const}}$ we have 
\be
\epsilon_{xy} (x,y)= -(1+\bar{\nu}) \frac{\partial \epsilon_{yy}(x,y)}{\partial y}  x.
\ee

Thus, this simple elasticity theory  allows us to determine the synthetic gauge fields  in the device using Eq.~(1),
\bea
A_x =-c (1+ \bar{\nu}) \epsilon_{yy}, ~~~
A_y =-2 c  \epsilon_{xy}.
\eea
where $c=\frac{t \beta}{e v_F}$.

Consider an adiabatically slow time-dependent Force of the form $F(t) = F_{DC}+F_{AC} \cos(\omega t)$. Using Eq.~(1) and the above relations we have
\bea
\label{eqsim}
B(x,y,t)&=&3c(1+\bar{\nu}) \partial_y \epsilon_{yy}=-3c (1+\bar{\nu}) \frac{ F}{h Y}\frac{\partial_y W(y)}{W(y)^2}, \nonumber \\
E_{x}(x,y,t)&=& c (1+\bar{\nu}) \partial_t \epsilon_{yy} = c (1+\bar{\nu}) \frac{\partial_t F}{h Y} \frac{1}{W(y)}, \nonumber \\
E_{y}(x,y,t)&=&2c \partial_t \epsilon_{xy}=2c(1+\bar{\nu}) \frac{\partial_t F}{h Y} \frac{\partial_y W(y)}{W(y)^2} x.
\eea

Relating the force to the stretch $\Delta y (t)=\Delta y + \sin(\omega t) \Delta y_{AC}$ using Eq.~(\ref{ForceStretch}), and here ignoring $\Delta y_{AC}$,  gives
\bea
B = \left( \frac{\Delta y}{L} \right) \left( \frac{1-f_r}{1+f_r} \right) \frac{1}{L} \times 6 c (1+\bar{\nu}).
\eea
The first, second, and third factors show the relation between the pseudomagnetic field and the relative stretch, the narrowing percentage, and the overall dimensions of the ribbon. The last dimensionfull factor can be estimated for graphene using $\frac{t}{ev_F} \approx 2.5 \mu$m T. For a relative stretch of $20 \%$, $f_r = 1/2$, and $L = 1\mu$m, as in Fig.~2, as well as $\beta \approx 2.5$ and $\bar{\nu}=0.17$, this estimate gives $3$ Tesla, which is close to our COMSOL simulation.

Similarly, the pseudoelectric field along $x$ reads
\be
E_{x}(y,t)=\left(\frac{\Delta y_{AC}}{L} \right) \omega \left(\frac{1}{1+f_r}\frac{W(L)}{W(y)}  \right) \times 4 (1+\bar{\nu}) c \cos(\omega t).
\ee
Note the $y$ dependence of the pseudoelectric field as given by $W(y)^{-1} \propto f_r (L-y)+y$ from Eq.~(\ref{shapefunc}), consistent with our COMSOL simulation in Fig.~2(b),(c).

\section{Appendix B: Suppressing intervalley scattering using smooth edges}

\begin{figure*}[ht]
	\includegraphics[scale=0.6]{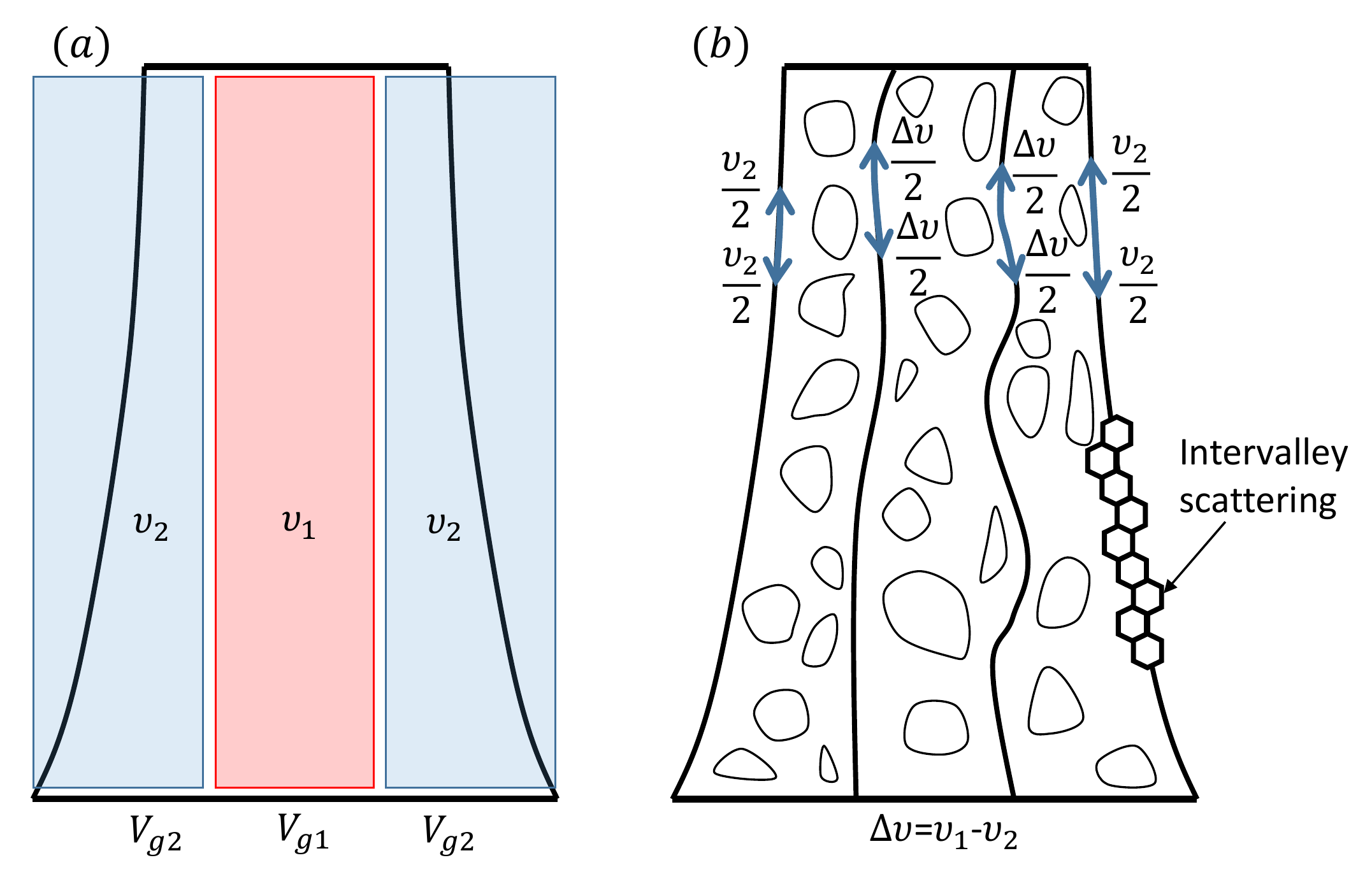}
	\caption{Device designed to suppress intervalley scattering at the effective edges. (a) Adding parallel gates allows one to define three parallel regions in the quantum Hall regime with different filling factors. (b) Schematic electron trajectories in the presence of the disorder potential, which is assumed to be smooth on the atomic scale. The exterior edge states are  sensitive to the graphene termination  which is expected to cause intervalley scattering, while the internal interface modes are only sensitive to the smooth potential and hence preserve the valley index. The total number of edge modes at each edge or interface are indicated, and are equally split among the two chiralities.}
	\label{fig3}
\end{figure*}

The pseudo QH effect described in this paper strongly relies on the absence of intervalley scattering. Intraedge intervalley relaxation will suppress the current by a factor $\sim e^{-L/L_{{\rm{iv}}}}$ where $L$ is the length of the edge and $L_{{\rm{iv}}}$ is the intervalley scattering length. 

In this appendix we argue that this assumption can be satisfied in the modified device in Fig.~5. The idea is that any sort of edge physics on the atomic scale, which typically contain an irregular combination of zigzag and armchair edges, produces  intervalley scattering. However  one can push the effective edges into the interior of the device, where their scattering becomes dominated by the disorder potential stabilizing
the QH effect~\cite{prange1990quantum,imry2002introduction} whose characteristic length scale is typically assumed to significantly exceed the atomic distance. As a result, the trajectories in the interior of the sample will have an approximately conserved valley quantum number. 

Consider adding three gates along the device as shown in Fig.~5(a), allowing independent control of the density  in the three  regions.  We envision these regions to stabilize separate gapped QH states, with the filling factor in the central region being $\nu_1$($ = \pm 2, \pm 6, \pm 10 \dots $), controlled by $V_{g1}$, and $\nu_2 \ne \nu_1 $ in the exterior regions which is assumed to be fixed and controlled by $V_{g
	2}$. 
In the bulk, the QH states consist of localized states, and the different QH states are separated by extended states, as shown in Fig.~5(b) where the typical disorder length exceeds the atomic scale. While the external edge mode trajectories between the $\nu_2$ region and vacuum are sensitive both to bulk disorder and to atomically sharp irregularities of the physical edge of the graphene sample, the interface edge modes between the $\nu_1$ and $\nu_2$ regions are determined solely by the smooth disorder potential. 

As a result of the smoothness of the disorder potential, the semiclassical trajectories corresponding to a superimposed fast cyclotron motion on top of a slow drift velocity $v_d = \frac{ \vec{B}  \times \nabla V_{\rm{dis}} }{|\vec{B}|^2}$ around the disorder potential, $   V_{\rm{dis}}$, have a well defined valley character. On the other hand, the edge modes near the edges of the sample are strongly affected by atomically sharp edge scattering and thus undergo intervalley scattering. These two types of 1D modes are spatially separated by the $\nu_2$  gapped QH region of localized states.

These filling factors dictate the number of edge modes at each interface.  For a real magnetic field the number of chiral edge modes  is given by the filling factor $\nu$, or by the difference of filling factors at an interface between two different QH states. But for a pseudomagnetic field these modes are equally split at each edge into the two chiralities, i.e., there are $\nu/2$ modes moving in each direction. As denoted in Fig.~5(b) the number of edge modes of each chirality is $\nu_2/2$ at the exterior edge and $|\nu_1 - \nu_2|/2$ at the interior interface between $\nu_1$ and $\nu_2$ filling factors. Let us assume $\nu_1 \ge \nu_2 > 0$ for simplicity.

Without intervalley scattering anywhere, the total two-terminal conductance as determined by the number of modes is dictated by the largest filling factor
\be
\frac{I}{V_{\rm{ext}}} |_{{\rm{no~iv~scattering}}} = \frac{e^2}{h}\nu_1
\ee
In the presence of strong intervalley scattering we assume that the external  edge modes of the $\nu_2$ region are gapped out and do not contribute. Then the two-terminal conductance becomes
\be
\frac{I}{V_{\rm{ext}}} |_{{\rm{strong~iv~scattering}}} = \frac{e^2}{h}(\nu_1-\nu_2).
\ee
As a function of the gate voltage $V_{g1}$ controlling $\nu_1$ the conductance will exhibit nearly quantized plateaus. The AC pseudo Hall effect will follow a similar behavior.

This analysis also implies that one can use such a device to probe the importance of intervalley scattering at the outer edge and test the length scale $L_{{\rm{iv}}}$.

\section{Appendix C: AC pseudo Hall current in the diffusive regime}
As the density is tuned through the extended PLL states, bulk transport takes place. This means that the pseudoelectric field leads to a finite valley current  perpendicular to the edges of the sample. Since the electric field is opposite for the two valleys, this leads to a valley polarization near the edges, which eventually in the DC limit leads to a diffusive current effectively screening the external pseudo-$E$ field and suppressing $I$. For finite frequency, this opposing diffusive current does not fully develop. In this appendix we present an approximate semiclassical analysis of the current at finite frequency. 

We consider a transport equation for the current densities  of the two valleys $\vec{j}_\pm$, which includes a  dissipative conductance $\sigma$, a diffusion current, and a  Hall effect, as well as the continuity equation,
\bea
\label{drude}\vec{j}_\pm &=& \pm \sigma \vec{E}(t) - D \vec{\nabla} n_\pm \mp (\omega_c \tau) \vec{j}_\pm \times \hat{z}, \\
\label{cont} 0 &=& \vec{\nabla} \cdot \vec{j}_\pm + \frac{dn_\pm}{dt}.
\eea
Here $D$ is the diffusion constant. % and $\mu$ is the mobility.
The currents $\vec{j}_\pm$ and the densities $n_\pm$ are related by the continuity Eq.~(\ref{cont}), and also satisfy boundary conditions $j_x(x=0,y)=j_x(x=W,y)=0$. Again here we ignore intervalley scattering and hence obtain  uncoupled equations for the two valleys.

We solve these equations under  simplifying assumptions of (i)  no $y-$dependence of neither the width $W(y) \to W$ nor the fields $E(y) \to E$, and (ii) the AC electric field  points along the $x$ direction only. Then the $y$ component of Eq.~(\ref{drude}) gives $j_y= \omega_c \tau j_x$, with $j_{x,y} \equiv (j_+)_{x,y}$, and the $x$ component of Eq.~(\ref{drude}) yields the differential equation 
\be
\label{eqLdrudediff}
[1+(\omega_c \tau)^2] j_x= \sigma E +\frac{D}{i \omega} \partial_x^2 j_x, 
\ee
with boundary condition $j_x(0)=j_x(W)=0$. The electric field is the real part of $E e^{i \omega t}$ and the current contains both in and out of phase components. From the diffusion time across the width of the sample
\be
\tau_T \equiv \frac{ W^2}{D},
\ee 
we form a dimensionless parameter $\omega \tau_T$ (which takes very low values in our system as estimated below). One recasts the differential equation in terms of dimensionless coefficients, a dimensionless variable $\tilde{x}=x/W$, and a source term,
\bea
j_y = \frac{\omega_c \tau}{1+ (\omega_c \tau)^2} \sigma E + \frac{1}{i (\omega \tau_T) (1+(\omega_c \tau)^2)} \nabla_{\tilde{x}}^2 j_y,
\eea
with boundary conditions $j_y(\tilde{x}=0)=j_y(\tilde{x}=1)=0$.
It is solved by
\bea
\label{solution}
j_y (x,t)= \frac{\sigma E (\omega_c \tau) }{1+(\omega_c \tau)^2} \times \tilde{j}(x/W)\nonumber \\
\tilde{j}(\tilde{x})= 1+ \left[ \frac{e^{-A}-1}{e^A-e^{-A} } e^{A \tilde{x}}+\left( A \to - A \right) \right],
\eea
where $A^2 = i (\omega \tau_T) [1+(\omega_c \tau)^2]$. Having solved for $j_+$ (valley $K$), we can obtain $\vec{j}_-$ by replacing $B \to - B$ and $E \to - E$.
Flow lines and the current profile as function of $\tilde{x}$ are plotted in Fig.~6.

At high frequency screening does not have time to develop. With $A \to \infty$ we have $j_y (x,t)= \frac{\sigma E (\omega_c \tau) }{1+(\omega_c \tau)^2}$, except right on the edge. This current is in phase with the electric field.

At low frequency we expect a strong suppression of the current due to screening. Expanding the solution for small $A$, corresponding to low frequency, we obtain 
\be
\label{lowomega}
j_y \xrightarrow[\omega \tau_T \ll 1]{\text{}} \frac{i}{2}  \sigma E  \cdot (\omega_c \tau)\cdot ( \omega \tau_T)\cdot \frac{x}{W} \left(\frac{x}{W} -1 \right),
\ee
which is out of phase with respect to the electric field. We see the  suppression factor $( \omega \tau_T)$ due to the scrreening effect, which becomes efficient when $\omega \tau_T \ll 1$.
Integrating over the width of the sample yields
\be
I=\int_0^W dx j_y(x)=\frac{i}{12} \sigma E W \cdot (\omega_c \tau) \cdot (\omega \tau_T) .
\ee

\begin{figure}[b]
	\centering
	\includegraphics[scale=0.4]{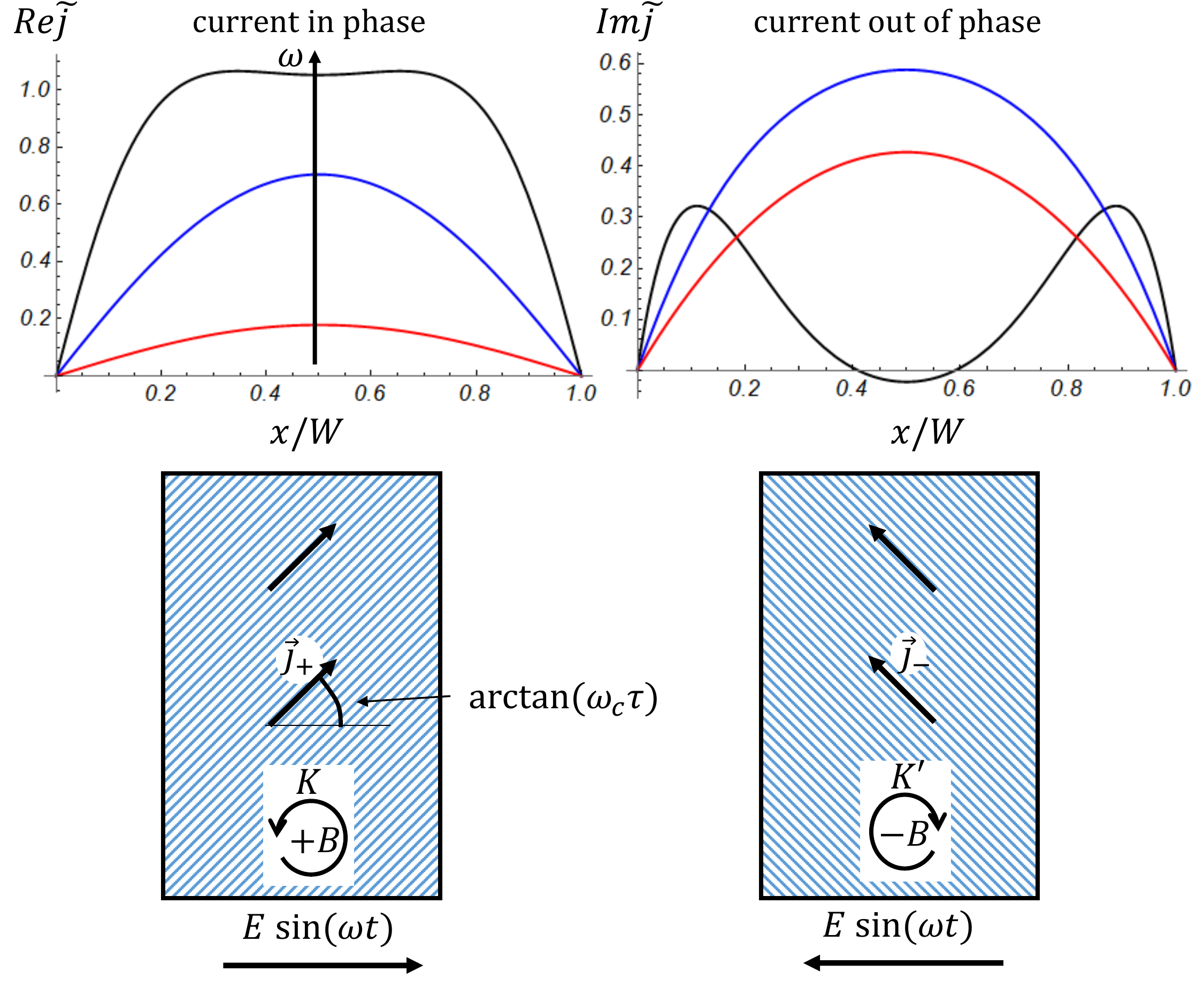}
	\caption{Plot of the current distribution Eq.~(\ref{solution}) in the diffusive regime as function of frequency $\omega$. The parameter $A \propto \omega$ takes the values $10,3.3,2$. In the high-frequency regime the current is in phase with the electric field, except near the boundaries, since screening is not effective. At low frequencies we obtain the parabolic current distribution in Eq.~(\ref{lowomega}) which is primarily out of phase.  }
	\label{fig3}
\end{figure}

We now estimate the diffusion time $\tau_T=W^2/D$ for a width $W \approx 0.5 \mu$m.  Doe the diffusion coefficient $D$ we use Einstein's relation $\sigma = D e^2 dn/dE_F$ with $\sigma \approx \frac{e^2}{h}$ a typical experimental value for the longitudinal conductivity in graphene at QH transitions~\cite{novoselov2005two,sarma2011electronic}. The density of states $dn/dE_F$ depends on disorder as depicted in the inset of Fig.~4. Here we estimate $dn/dE_F$ very crudely by assuming a density of states of a clean LL, $\frac{B}{\Phi_0} \delta(E- E_{LL})$, spread due to disorder over an energy approximately given by the LL spacing $\hbar \omega_c =v_F \sqrt{\hbar e B}$. This estimate is equivalent as an order of magnitude to the Dirac density of states $dn/dE_F=\frac{2}{ \pi} \frac{k_F}{\hbar v_F}$ with $k_F$ determined from the electronic density of a full LL,  $k_F^2 \sim \frac{B}{\Phi_0}$. For $B=3$T this gives $k_F \sim 10^{2} \mu {\rm{m}}^{-1}$ and we obtain $D \sim 0.01 {\rm{m}}^2 s^{-1} $ giving a time of $\tau_T \approx 10^{-10}~{\rm{s}}$. For a typical  piezoelectric-mechanical frequency $\omega \approx 10^7 {\rm{Hz}}$ we have $\omega \tau_T  \sim 10^{-3}$.

While here we considered piezoelectric modulators, which sets an upper limit for possible frequencies, the effect  can in principle be observed at higher frequencies.
\end{document}